      [1999/12/01 v1.4c Il Nuovo Cimento]
\documentclass{cimento}

\usepackage{graphicx}  
\title{The Fermi Gamma--ray Burst Monitor: \\
Results from the first two years}
\author{E.~Bissaldi\from{ins1}\from{ins2}\ETC,
on behalf of the Fermi/GBM Collaboration}
\instlist{\inst{ins1} Max--Planck--Institut f\"ur extraterrestrische Physik, Giessenbachstr.~1, \\ 85748 Garching, Germany
          \inst{ins2} Institute of Astro and Particle Physics, Technikerstr.~25, 6020 Innsbruck, Austria}
\PACSes{\PACSit{95.55.Ka}{X-- and $\gamma$--ray telescopes and instrumentation} 
\PACSit{98.70.Rz}{$\gamma$--ray sources; $\gamma$--ray bursts}}
\begin{document}

\maketitle

\begin{abstract}
In the first two years since the launch of the Fermi Observatory,
the Gamma--ray Burst Monitor (GBM) has detected over 500 Gamma--Ray Bursts
(GRBs), of which 18 were confidently detected by the Large Area Telescope
(LAT) above 100 MeV. Besides GRBs, GBM has triggered on other transient sources, such as Soft
Gamma Repeaters (SGRs), Terrestrial Gamma--ray Flashes (TGFs) and solar
flares. Here we present the science highlights of the GBM observations.
\end{abstract}
\section{Introduction}
The Fermi Gamma--ray Space Telescope,
which was successfully launched on June 11, 2008, is an international 
and multi--agency space observatory. The payload comprises two science
instruments, the Large Area Telescope
\cite{ATW09}, a pair conversion telescope operating 
in the energy range between 20
MeV and 300 GeV and the Gamma--Ray Burst Monitor 
\cite{MEE09},
which extends the Fermi energy range to lower energies 
(from 8 keV to 40 MeV).
The primary role of the GBM is to augment the science
return from Fermi in the study of GRBs by discovering 
transient events within a larger field of view (FoV) and 
performing time--resolved spectroscopy of the measured 
burst emission.
\section{The GBM Detectors}
The GBM flight hardware comprises 12 thallium activated
sodium iodide (NaI(Tl), hereafter NaI) scintillation detectors and two bismuth
germanate (BGO) scintillation detectors.
The individual  NaI detectors ($\sim$8--1000 keV) are mounted around the 
spacecraft and their axes are oriented such that the
positions of GRBs can be derived from the measured relative
counting rates. With their energy range extending between $\sim$0.2 and 
$\sim$40~MeV, two BGO detectors provide the overlap in 
energy with the LAT instrument and are crucial 
for in--flight inter--instrument calibration \cite{BIS09}. 
They are mounted on opposite sides of the Fermi spacecraft 
covering a net $\sim$8 sr FoV.

A burst trigger occurs when the flight software detects an
increase in the count rates of two or more detectors above an
adjustable threshold specified in units of the standard deviation
of the background rate. 28 different
trigger algorithms operate simultaneously, each with a distinct threshold.
The trigger algorithms currently implemented include four
energy ranges: 50--300 keV, which is the standard GRB trigger range, 25--50 keV 
to increase sensitivity for SGRs and GRBs with soft
spectra, $>$100 keV, and $>$300 keV to increase sensitivity for
hard GRBs and TGFs. 
When a burst trigger occurs, onboard software determines a
direction to the source using the relative rates in the 12 NaI
detectors. These rates are compared to a table of calculated
relative rates. The location with the best chi--squared
fit is converted into right ascension and declination
using spacecraft attitude information and transmitted to the
ground. Improved locations are automatically computed on the ground
in near real--time by the Burst Alert Processor (BAP)
and later interactively by the GBM Burst Advocate.

The Fermi Observatory incorporates the capability to autonomously
alter the observing plan to slew to and maintain
pointing at a GRB for a set period of time, nominally 5 hours, subject
to Earth limb constraints. This allows the LAT to observe
delayed high--energy emission, as has been previously observed
by instruments on the Compton Gamma--Ray Observatory \cite{HUR94}. Either the
GBM or the LAT can generate an Autonomous Repoint Request
(ARR) to point at a GRB. A request originating from
GBM is transmitted to the LAT. The LAT either revises the
recommendation or forwards the request to the spacecraft. The
LAT software may, for example, provide a better location to the
spacecraft, or cancel the request due to operational constraints.
The GBM flight software specifies different repoint criteria
depending on whether or not the burst is already within the LAT
FoV, defined as within 70$^o$ of the +Z axis. An ARR is generated
by GBM if the trigger exceeds a specified threshold for peak flux
or fluence. These thresholds are reduced if the burst spectrum
exceeds a specified hardness ratio.
\section{Two years scientific results}
\subsection{GRB observations}
During its two years of operations since 
trigger enabling (July 13th, 2009 -- September 6th, 2010), GBM detected 
$\sim$540 GRBs, with $\sim$270 bursts placed inside the 
LAT FoV. In 45 cases, an ARR was positively issued by GBM to repoint the
LAT and follow particularly bright events. Morover, 18 GRBs 
were detected at energies above 100 MeV. The most
energetic events include GRB~080916C \cite{ABD09a},
GRB~090510 \cite{ABD09b},
GRB~090902B \cite{ABD09c}, and
GRB~090926A  \cite{ABD10}. In all four cases, the LAT detected
photons with energies $>10$ GeV, with GRB~090902B being the record holder
with 33 GeV, the highest photon energy reported so far for a GRB.
Moreover, GRB 090510 was the first short GRB showing emission 
up to 31 GeV. For this burst, no evidence was found for the
violation of Lorentz invariance, thus disfavoring quantum--gravity 
theories in which the quantum nature of spacetime on a very 
small scale alters the speed of light in a way that
depends linearly on photon energy.
From the spectral point--of--view, joint GBM--LAT observations
uncovered the presence of two spectral components in GRB~090902B. 
For this burst, a power law component
dominates the Band component at both low ($<50$
keV) and high ($>100$ MeV) energies and is significant within 
just the GBM data.  Furthermore, GRB~090926A showed
evidence for a high energy cutoff.

Recently, \cite{GUI10} presented an extremely detailed time--resolved 
spectroscopy at timescales as short as 2 ms for the three 
brightest short GRBs observed with GBM (GRB~090227B, GRB~090228 
and GRB~090510). The time--integrated spectra of the 
events deviate from the Band function, indicating the existence 
of an additional spectral component. The time--integrated Epeak 
values exceed 2 MeV for two of the bursts, and are well above 
the values observed in the brightest long GRBs. Their Epeak 
values and their low--energy power--law indices confirm 
that short GRBs are harder than long ones. We find that 
short GRBs are very similar to long ones, but with light 
curves contracted in time and with harder spectra stretched 
towards higher energies.

\begin{figure}[t!]
\centering
\includegraphics[width=0.36\textwidth,bb=70 2 560 790,clip,angle=90]{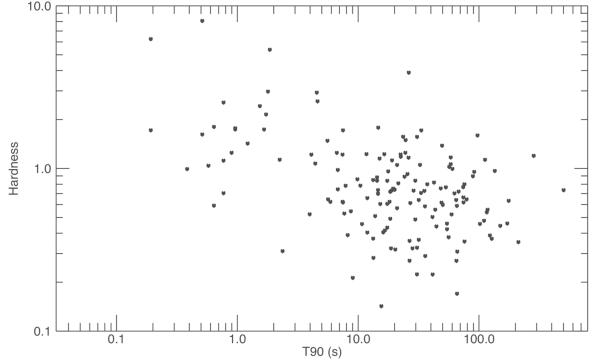}
\caption{
Distribution of spectral hardness versus duration 
for a portion of the GBM bursts
}
\label{fig1}
\end{figure}

Regarding common Fermi--{\it Swift} \cite{GEH04} GRB observations,
the overlapping FoVs of GBM and of the Burst Alert Telescope (BAT)
onboard {\it Swift} contributed to $\sim$40 common detections/yr
and resulted in 1 common GBM/LAT/BAT detection so far, namely 
in the case of the long--lived emission (from UV to GeV) of GRB~090510
\cite{DEP10}. The finer LAT localizations errors ($<0.4^{\rm o}$)
resulted in Swift follow--up of several Fermi bursts,
providing 8 redshifts measurements (by ground--based observatories) so far.
\subsection{GRB Catalog}
Catalogs covering the first two years of GBM GRBs are currently 
being produced. The main catalog \cite{PAC11}
summarizes basic parameters for all triggered GRBs, including sky location, 
fluence in two energy ranges (50--300 keV and 10--1000 keV), peak flux for 
the same two energy ranges and three timscales (64 ms, 256 ms and 1024 ms)
and duration (T50 and T90 in the 50--300 keV range). The spectroscopy 
catalog \cite{GOL11} includes spectral 
fits using several standard functions for all sufficiently bright GRBs. 
The standard functions include power--law, power--law with exponential 
cut--off, Band function, and smoothly--broken power--law. For each 
burst two sets of spectral fits are performed: a 3.5 $\sigma$ 
signal--to--background selection for the duration of the burst 
(fluence spectra), and a peak count rate selection (peak flux spectra). 
The peak count rate interval is 1 s for long bursts and 64 ms for 
short bursts, based on the burst T90 duration.
Fig.~\ref{fig1} shows the distribution of spectral hardness versus 
duration for a portion of the GBM bursts. Although the results are
preliminary, the well--known tendency for short bursts to have harder 
spectra is clearly shown.
\subsection{Non--GRB science}
Apart from GRBs, GBM triggered 170 times on SGR outbursts,
mostly on soft, short trigger algorithms. Those triggers originated from four
different active SGRs. Of those four, one is an already known source (SGR J1806--20),
two are new sources discovered with {\it Swift} (SGRJ 0501+4516 and SGR J1550--5418) and one
was discovered with GBM (SGR J0418+5729) \cite{VAN10}.
An interesting finding by \cite{KAN10} was the identification of a 
$\sim$150--s--long enhanced persistent emission during
the January 2009 outburst of SGR J1550--5418, which exhibited intriguing 
timing and spectral properties. The area of the blackbody emitting
region which could be estimated ($\approx$0.046 ($D$/5 kpc)$^2$ km$^2$,
{\it i.e.} roughly a few $\times10^5$ of the neutron star
area) is the smallest ``hot spot'' ever measured for a magnetar and most likely
corresponds to the size of magnetically--confined plasma near the neutron
star surface.

GBM also triggerd 93 times from TGF signals, all on hard, short trigger algorithms
(see also \cite{WIL11}).
First observations of a smaller sample of TGFs and their properties
are reported in \cite{BRI10}. The temporal properties 
of a considerably larger sample of TGFs observed with GBM can be found in
\cite{FIS11}, who discuss several distinct categories of TGFs mainly identified by
their time profiles. Moreover, \cite{CON10} recently 
presented a search for correlations between TGFs 
detected by GBM and lightning strokes measured using 
the World Wide Lightning Location Network (WWLLN) \cite{ROD09}.
In addition to gamma--ray TGFs, GBM has observed several TGFs by the
propagation of charged particles along geomagnetic field lines. Strong 511
keV annihilation lines have been observed, demonstrating that both electrons
and positrons are present in the particle beams \cite{BRI11}.

Finally, GBM triggered 33 times on solar flares. Several other triggers 
were caused by well known gamma--ray emitters, such as Cyg X--1, or by 
accidental particle events.
Besides triggered transient sources, which are detected by the 
on--board trigger algorithm, GBM can be used to study
hard X--ray pulsars with
periods greater than a few seconds. These are monitored using
Fourier transforms and epoch folding \cite{FIN10}.
Moreover, GBM background data is extremely useful for a number of other studies, 
enabling a wide range of guest investigations. 
These data are currently used to monitor variable
X--ray sources using the Earth occultation technique \cite{WIL10}.

\end{document}